\documentstyle[osa,prb,twocolumn,aps,graphicx]{revtex}

 \newcommand{\etal}{\mbox{\it et al.}}
   \newcommand{\beq}{\begin{equation}}
   \newcommand{\eeq}{\end{equation}}
   \newcommand{\bea}{\begin{eqnarray}}
   \newcommand{\eea}{\end{eqnarray}}
   \newcommand{\beastar}{\begin{eqnarray*}}
   \newcommand{\eeastar}{\end{eqnarray*}}

   \newcommand{\dxy}{d_{x^{2} - y^{2}}}

   \newcommand{\bk}{{\bf k}}

         \newcommand{\cd}{c^{\scriptscriptstyle \dag}}
         \newcommand{\sss}{\scriptscriptstyle}
         \newcommand{\cid}{c^{{\sss \dag}}_{i\sigma}}

         \newcommand{\cj}{c^{}_{j\sigma}}

   \newcommand{\bdm}{\begin{displaymath}}
   \newcommand{\edm}{\end{displaymath}}

  \newcommand{\journal}[5]{#1, \textit{#2} \textbf{#3}, #4 (#5).}
         \newcommand{\jpc}{J. Phys.: Condens. Matter}

   \newcommand{\book}[3]{#1, \textit{#2} #3.}

\begin{document}
\draft
\twocolumn[\hsize\textwidth\columnwidth\hsize\csname
@twocolumnfalse\endcsname

\title{Groundstate properties of the BCS-Bose Einstein crossover in a 
$d_{x^{2}-y^{2}}$ wave superconductor} 
\author{B.C. den Hertog\cite{newaddress}} 
\address{Department of Physics, University of Waterloo, ON Canada N2L 3G1 \\
and\\
Department of Theoretical Physics,
 Institute of Advanced Studies, The Australian National University,
 Canberra ACT  0200, Australia}

\maketitle
\begin{abstract}
 We examine the groundstate properties of the  crossover between  BCS 
superconductivity and Bose Einstein condensation    
within a  model which exhibits $d_{x^{2} - y^{2}}$ pairing symmetry. 
We compare results for zero temperature with known features of $s$-wave 
systems and show that 
bosonic degrees of freedom are likely to emerge  only in the dilute limit. 
We relate the 
change in properties of the groundstate as a function of density 
 to the effect of the exclusion principle and
show that the one particle distribution function undergoes a significant 
change when  bosonic behavior appears.
\end{abstract}

]
\section{Introduction}
The existence of a crossover from BCS superconductivity to Bose Einstein (BE) 
condensation  of pre-formed pairs has received  increasing 
attention in the literature over the last few years. Originally discussed by 
Eagles \cite{Eagles} within the context of pairing in thin film 
semi-conductors, and considerably expanded upon  by the works of Leggett 
\cite{Leggett} and Nozi\`{e}res and Schmitt-Rink \cite{Nozieres3}, the latest 
resurgence in interest has been due to its possible application to the 
understanding of the phase diagram of high temperature cuprate
 superconductors. 

 In
 particular, the appearance of a pseudo-gap in both charge and spin 
excitations 
 of the normal state \cite{Cooper,Batlogg,Takigawa,Marshall,Loeser} 
has led  to  a number of  suggestions  that  pairing correlations well 
above 
$T_{c}$ may occur in these systems. Whether these correlations
are in the bosonic form of pre-formed pairs 
\cite{Ranninger,Alexandrov},  or pair 
resonances originating from the intermediate regime of a BCS-BE crossover\cite{Janko,Trivedi},  or simply from classical phase 
fluctuations\cite{Kivelson} is still however a controversial issue.

 Furthermore, now that there 
is  a large body of evidence to suggest that pairing in the cuprates is 
predominantly $\dxy$ in character \cite{Ding,Tsuei,Annett},  there is  
a clear need to  understand the properties of the BCS-BE   crossover 
within the context of this type of pairing symmetry. 
Although there have been some 
attempts to discuss the effect of $\dxy$ pairing on pseudo-gap formation above $T_{c}$ in the cuprates\cite{Engelbrecht}, there has been little discussion on the systematic 
groundstate properties of the BCS-BE crossover in the $\dxy$ channel.
 This is an important issue, since pairs with 
this symmetry cannot contract to point like bosons and so accordingly one expects that there will be severe 
consequences for the properties of the groundstate of this type of  system. This is of 
 direct importance in terms of the validity of the crossover scenario for 
the cuprates, whilst also being of general interest in understanding 
macroscopic  pairing in higher angular momentum channels.

In this paper, we consider the BCS-BE crossover at zero temperature as a 
function of both coupling strength and carrier density, 
within a two-dimensional toy model which has 
a $\dxy$ pairing instability. The study  at zero temperature is well 
controlled by use of the BCS variational 
wavefunction (which contains the BE limit \cite{Leggett}) and allows us to 
establish at the two body level  the qualitative groundstate properties of the system 
upon which future  studies may be based. In particular we show that the 
groundstate 
properties of the system are severely modified from the $s$-wave picture, 
where a smooth crossover exists for all densities \cite{Leggett,Nozieres3}.  
For the $\dxy$ case  
the effect of the exclusion
 principle as the density of carriers is increased  results in a  suppression 
of the emergence of bosonic degrees of freedom for moderately 
large densities and strong coupling, where the system instead remains 
fermionic (BCS like). We find that only in the dilute limit is a crossover  
possible, and that when this occurs 
\begin{center}
\begin{figure}
\includegraphics[width=8.0cm]{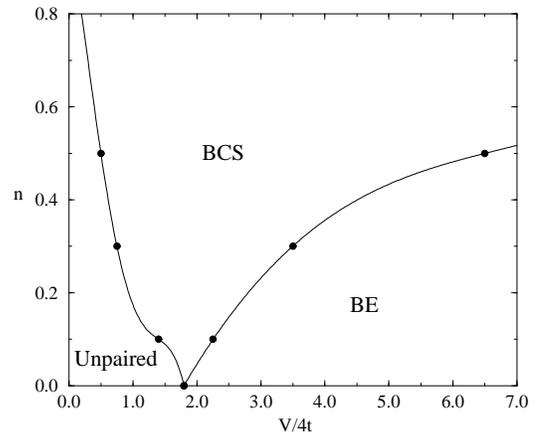}
\caption{The crossover between fermionic and bosonic degrees 
of freedom as a function of carrier density and coupling strength (as defined
 in text). The left boundary is determined by the onset of a finite gap amplitude, and the right boundary by when the chemical potential falls below the band minimum. The solid lines are a guide for the eye only.}
\label{bec-phase}
\end{figure}
\end{center}
the single particle distribution function  undergoes a radical change. 
We summarize most of our results in Fig. \ref{bec-phase}, which shows  when  bosonic degrees of freedom can emerge as  a function 
 of  both density and coupling strength.

\section{Toy Model}
 We introduce as our  
toy model a `reduced'  Hamiltonian in  the  BCS  sense  which
 describes an 
effective two-particle interaction in real space  and 
in  the singlet pairing channel \cite{note2}. Specifically it  is 
 \begin{eqnarray}
 \label{hamil4}
 H&=&\sum_{<ij>\sigma}-t(\cid\cj + h.c.) -\mu^{*}\sum_{i\sigma} n_{i\sigma} 
 \nonumber \\
&+&
  W\sum_{i} c^{{\sss \dag}}_{i\uparrow}c^{{\sss \dag}}_{i\downarrow} 
 c^{}_{i\downarrow}c^{}_{i\uparrow} 
- 
 V \sum_{<ij>}c^{{\sss  \dag}}_{i\uparrow}c^{{\sss 
 \dag}}_{j\downarrow}c^{}_{j\downarrow}c^{}_{i\uparrow} , 
 \end{eqnarray}
 where first and second  terms describe  nearest neighbor hopping on a 
 two-dimensional square lattice with chemical potential $\mu^{*}$, and the 
third and fourth terms 
 describe an effective two-body pairing interaction in real space. In 
particular, 
 the third term 
 represents the repulsive part of the effective interaction  while 
the last term provides an attractive 
 interaction for nearest neighbor particles\cite{note10}.

By expressing the superconducting  gap in terms of its various symmetry
 components and  
introducing the BCS variational wavefunction 
$|\Psi\rangle=\prod_{\bk}(u_{\bk} + 
v_{\bk}\cd_{\bk\uparrow}\cd_{-\bk\downarrow})|0\rangle$, the 
zero temperature gap equation in the  $d_{x^{2}-y^{2}}$ channel is given by, 
\begin{equation}
\label{gap1} 
 1=\frac{1}{2M}\sum_{\bk}\frac{V(\cos k_{x} - \cos 
 k_{y})^{2}}{\left[(\xi(\bk) - \mu)^{2} + 
 \Delta^{2} (\cos k_{x} - \cos k_{y})^{2}\right]^{1/2}} ,
\end{equation}
where $\xi(\bk)=-2t\eta_{k}$ is the nearest 
neighbor tight-binding energy dispersion, the geometric factor 
$\eta_{\bk}=\cos k_{x} + \cos k_{y}$ and the effective chemical potential 
$\mu=\mu^{*} -n(W/2 -2V)$.  The amplitude of the d-wave gap is
 denoted by $\Delta$. 

As Eagles first pointed out \cite{Eagles}, any deviation from weak 
coupling requires  a self consistent solution of both  gap and number 
equations, since the BCS approximation of the chemical potential being equal 
to its value in the normal state can no longer  be reasonably justified.
 Specifically, the number equation
 for  the chemical potential $\mu$, which defines the  particle density 
$n=N/M$ is ,  
 \beq
 \label{num1}
 n-1=\frac{1}{M}\sum_{\bk}\frac{-(\xi(\bk) - \mu)}
{\left[(\xi(\bk)-\mu)^{2} + 
 \Delta^{2}(\cos k_{x} - \cos k_{y})^{2}\right]^{1/2}} .
 \eeq

 We have solved Eqns. (\ref{gap1}) and (\ref{num1}) self consistently 
 at a given density by numerical integration.  
In this theory, pairing takes place over 
 the entire Brillouin zone (BZ).  The  zone edge acts 
as a natural  boundary or momentum cut-off  which avoids the renormalization 
methods  used to remove ultraviolet divergences in
  continuum model treatments of strong coupling within the 
 BCS framework \cite{Randeria4}. 
\section{Crossover and Analysis}
\begin{center}
\begin{figure}
\includegraphics[height=6.5cm,width=8.5cm]{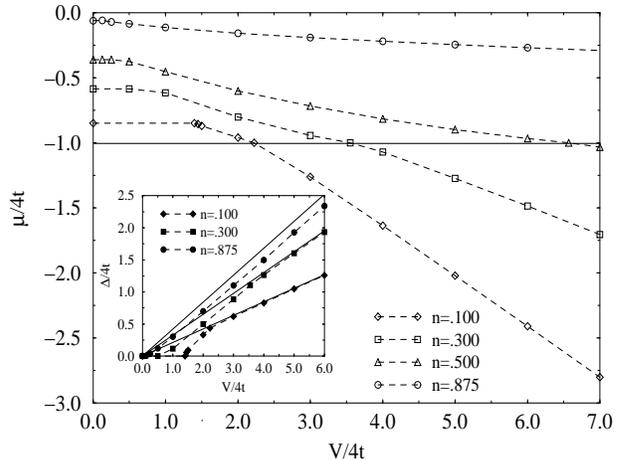}
\caption{The dependence of  $\mu$ on  
coupling strength for various  densities. 
{\it Inset:} The gap  $\Delta$ as a function of  coupling 
strength (dashed 
lines and data points). The solid 
lines are  the asymptotic behavior derived in the 
text.}
\label{gapdata}
\end{figure}
\end{center}

The inset of  Fig.  \ref{gapdata} 
shows the gap parameter 
 $\Delta$ for densities ranging from 
 the dilute limit to the almost half-filled case.
 Taking the onset of 
 a finite gap parameter  as the signal for the 
 manifestation of the superconducting state,
 an 
 increase in the particle density clearly favors  the emergence of 
 a $\dxy$ paired groundstate. 
On the other hand, as the system becomes more 
dilute, there is a  
 need for stronger coupling to induce pairing. 
 
In the  dilute limit (of relevance to the underdoped cuprates, where the 
 density of carriers is proportional to the doped hole  concentration  $x$), 
the occurrence of  $\dxy$  
 bound states for the two-particle problem on an empty 
 lattice is for our system,  equivalent to that problem for the $t-J$ 
 model.  Kagan and Rice \cite{Kagan} have shown that a $\dxy$ 
bound  state will not occur unless the coupling $J$ (equivalent to $V$ in our 
case) is greater  than  $V_{c}/4t \sim 1.8$. 
Also,  Randeria \etal \cite{Randeria6} have shown that in two dimensions, a
necessary and sufficient condition for a dilute many-body $s$-wave  Cooper 
instability to  occur  is that an $s$-wave  
bound 
state exists for the corresponding two-body problem on the empty 
lattice. Importantly, in the context of   $\dxy$ symmetry, 
they have also shown that such 
a  condition does not exist for higher angular momentum pairing. 

In our study, we find that the onset of pairing occurs for progressively 
weaker coupling  as the density is increased, contrary to an $s$-wave system.
In the $\dxy$ channel at a coupling strength less than $V_{c}/4t$, the dilute 
system  gains more kinetic energy compared to pairing energy  and thus has a 
total condensation energy which is positive, whereas for the more dense system
 the kinetic energy of the carriers is on average less effected and a pairing 
instability occurs. This  steady evolution from the 
extreme dilute limit where pairing does not occur until $V/4t \geq
1.8$, to the near half filled case where superconductivity can 
manifest itself at a much weaker coupling than that required for a two body
bound state, is indicated by the boundary on the left in Fig \ref{bec-phase}. 
The $n=0$ point corresponds to the critical coupling strength $V_{c}$ 
discussed above and calculated initially by  Kagan and Rice\cite{Kagan}.

In  Fig. \ref{gapdata}  we show $\mu$ as found in 
conjunction with the
solutions for the gap $\Delta$ at various densities and as a function of the 
coupling 
strength. The horizontal line at $\mu/4t=-1.0$ represents the bottom of the tight 
binding band.  For weak coupling, it is well known that $\mu$ in the
 superconducting phase is given roughly by the Fermi energy of the normal 
state and this can be seen in the figure. However, at large doping,
$\mu$ shows little deviation from its 
normal state value over a large variation in the the coupling strength. 
This is to be contrasted with the low density results, which show a relatively
 rapid deviation from weak coupling behavior   as $V$ is increased. 

In general, bosonic degrees of freedom can be expected to emerge once 
the chemical potential of the many-body groundstate  slips below the band 
minimum in a tight binding system, or below zero in a continuum model. For an 
$s$-wave system,  Nozi\`{e}res and  Schmitt-Rink \cite{Nozieres3} were able to
 show that a crossover from fermionic superconductivity to bosonic degrees o freedom can occur for
 all densities as the coupling strength is increased. For the $d$-wave system
 considered here, this is not the case. Bosonic degrees of freedom can only 
emerge in the dilute regime, while for large densities, the system behaves more
 like a weak coupling superconductor with a value of the chemical potential 
comparable to that of the normal state. These results are expressed by the 
 boundary on the right  in Fig. \ref{bec-phase}. Given that the BCS wavefunction only takes two-body correlations into account, we expect that the suppression of bosonic degrees of freedom would be increased in a more sophisticated treatment which would include a repulsive interaction between fermion pairs\cite{Haussmann}.

Further insight into this feature of a $d$-wave system can be gained by 
examining the limit of infinite coupling strength. 
In the   case $V\rightarrow\infty$,  the kinetic term (and thus any Fermi surface geometry) becomes negligible and 
the asymptotic behavior of 
the gap and chemical potential can be shown from Eqns. (\ref{gap1}) and (\ref{num1}) to have the form
$\Delta\rightarrow\gamma V/2$
and
$\mu\rightarrow\gamma V (n-1)/2\alpha$, implying that in the infinite coupling
 limit, $\mu/\Delta\rightarrow(n-1)/\alpha$. Here   the parameter $\alpha$ is
 given by the solution to 
\beq
\alpha=\frac{1}{M}\sum_{\bk}\frac{1}{\left[ (\cos k_{x} - \cos 
k_{y})^{2} +
 ([n-1]/\alpha)^{2} \right]^{1/2}} ,
\eeq
while $\gamma$ is defined by
\beq
\label{gammastrong}
\gamma=\frac{1}{M}\sum_{\bk}\frac{(\cos k_{x} - \cos 
k_{y})^{2}}{\left[ (\cos k_{x} - \cos k_{y})^{2} + ([n-1]/\alpha)^{2} 
\right]^{1/2}} .
\eeq

We have indicated the asymptotic behavior of $\Delta$ at various densities in 
the inset  of
Fig. \ref{gapdata} by the solid lines. As one increases the 
density, the 
convergence to the asymptotic behavior is poor for large densities, indicating
that the $\mu$ is still of the order of the band energy.
 In the   strong coupling regime, if
true   bosonic characteristics emerge then  one expects that $\mu$ 
should simply reduce to the binding energy per 
particle for
 the `diatomic' electron molecule. It can  readily be  shown that only in the 
dilute limit  does the ratio 
$\gamma/\alpha$ approach unity. This indeed leads to the result $\mu=-V/2$, 
exactly one half of the binding energy for the two  body problem in the strong
coupling 
limit \cite{note3}.
  On the other hand,
 at half filling  ($n=1$)  $\mu$ 
clearly remains zero even in the limit of infinite coupling.  Remarkably, the 
groundstate remains a  fermionic  superconductor for all coupling   strengths and 
has
characteristics equivalent to a weak coupling BCS system. For 
densities in 
between these two limits, an increase in coupling  will eventually 
reduce the
 chemical potential below the bottom of the band, however it  will 
always be 
somewhat greater than the binding energy per particle of the 
two-body case. 

This general dependence on the density for the BCS-BE crossover in $d$-wave systems
can be viewed as a manifestation of the effect  of the exclusion principle. 
Due to the symmetry of the pairs, they cannot contract in real space to point 
bosons, 
but must always retain a finite spatial extent. As the density 
increases, the overlap of the pair wavefunctions exerts its influence 
through the exclusion principle contributing a positive energy to the 
system. 
At half filling, this overlap  prevents the system from crossing over 
to a 
system displaying bosonic qualities  even in the infinite coupling 
limit.

This interpretation can be  further supported by a calculation of the 
average pair 
coherence length. We can define this length $\xi_{0}$ through the 
expectation value of the quantity 
$
\xi_{0}^{2}=\langle F_{\bk}|-\nabla^{2}_{\bk}|F_{\bk}\rangle/\langle F_{\bk}|F_{\bk}\rangle$ 
where $F_{\bk}=u_{\bk}^{*}v_{\bk}$ plays the role of the pair wavefunction \cite{Leggett2}.

\begin{center}
\begin{figure}
\includegraphics[width=8cm]{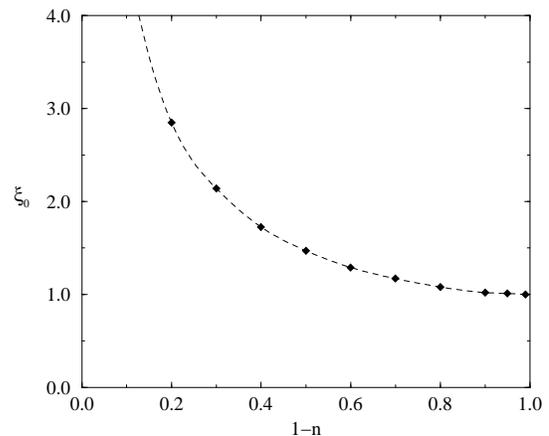}
\caption{The coherence length in the strong coupling limit as 
a function of  density. The lattice spacing has been 
set to unity.}
\label{corrlength}
\end{figure}
\end{center}

In Fig. \ref{corrlength} we show the behavior of $\xi_{0}$ in 
the 
infinite coupling limit as a function of density. It is clear  
that a shrinking of the pair size in  real space to a compactified  
boson spread out
 over   nearest neighbor sites can only 
 occur  in the  dilute limit, where the exclusion principle 
becomes irrelevant. For higher density, the overlap of the pair wavefunctions
  has the effect of increasing the correlation between   particles, which then
  increases the average 
pair size  in order that the  total energy of the system can be minimized. 

Lastly, we examine the single particle distribution function 
$n_{\bk\sigma}$ 
as a function of the coupling strength. We find that in weak coupling, 
the 
system behaves as a  typical BCS system, with $n_{\bk\sigma}$ 
resembling
 a step function up to a point close to the normal state Fermi 
surface, where
 upon it becomes smeared over a small region about $\mu=E_{F}$. The 
degree of
 smearing increases with the coupling until the chemical potential 
drops below the bottom of the band. At this point, the behavior of 
the single 
particle distribution function radically changes. 

\begin{center}
\begin{figure}
\includegraphics[height=5cm]{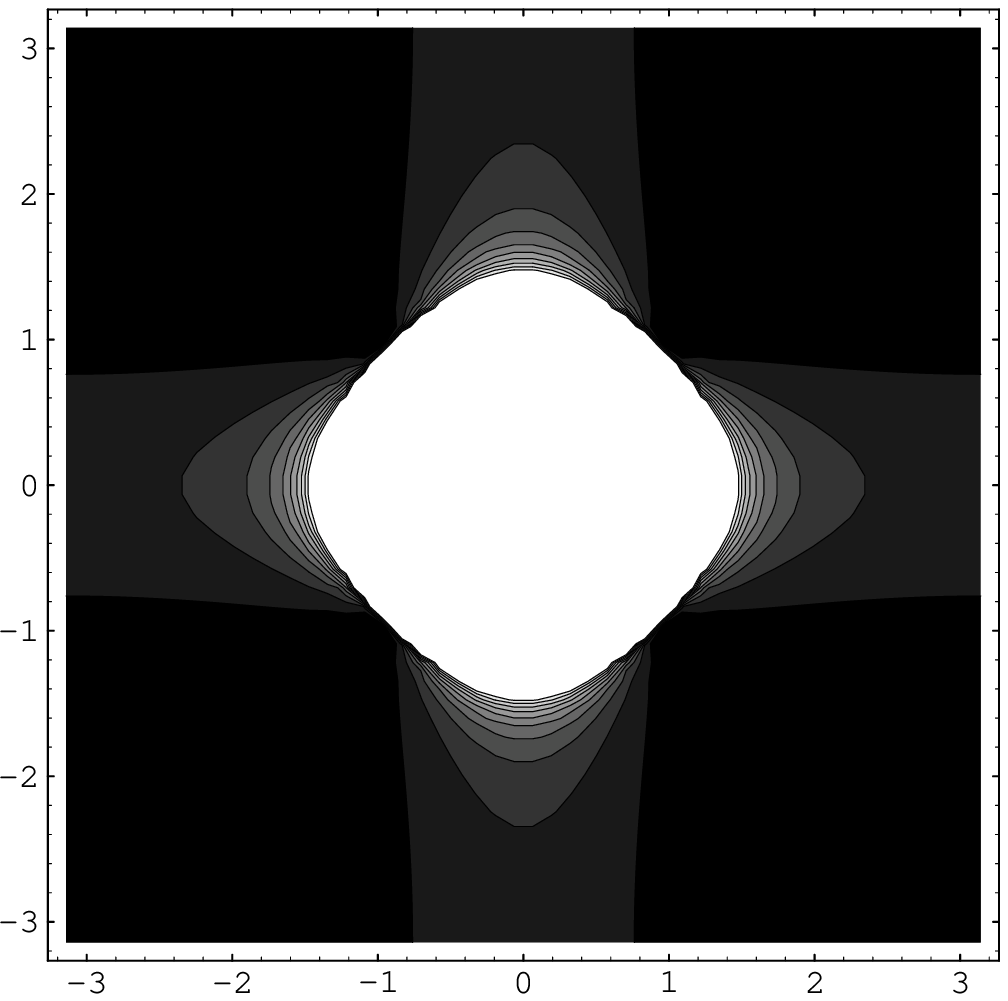}
\includegraphics[height=5cm]{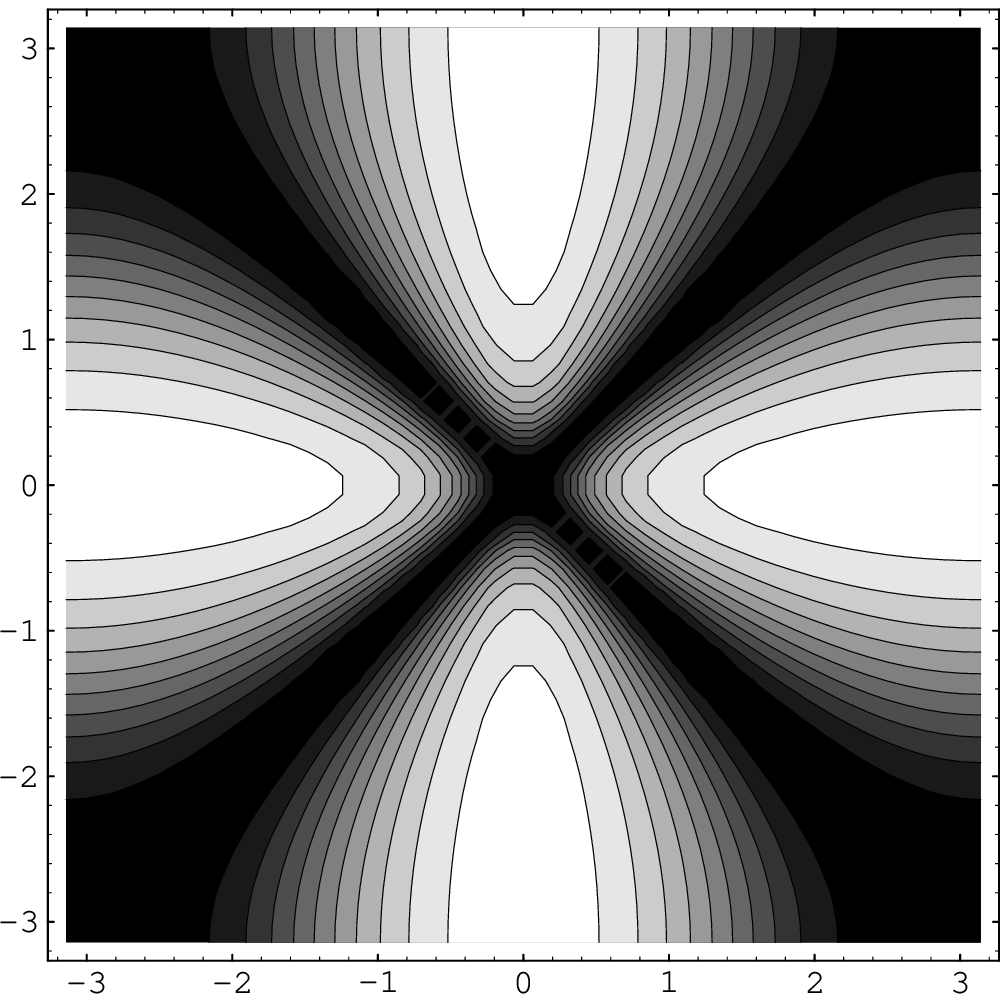}
\caption{Contour plots within the first BZ  of  $n_{\bk\sigma}$ for 
$V/4t=1$ (top) and $V/4t=4$ (bottom)  and  $n=0.3$. The brighter the region the larger the value of $n_{\bk\sigma}$. }
\label{occupation}
\end{figure}
\end{center}

In Fig. \ref{occupation}, we 
compare  
 contour plots of $n_{\bk\sigma}$ for $n=0.3$ throughout the  
BZ  for the weak coupling case $V/4t=1$, where $\mu$ is well 
approximated by its normal state value,  and the stronger coupling case $V/4t=4$ 
where the 
chemical potential falls just below the bottom of the band.

In the 
weak 
coupling case, the Fermi surface of a tight binding band with a small  density 
per unit volume can 
be clearly  seen by the bright region in the middle of the plot. 
On the other hand, when $\mu$ drops below the bottom of the band and bosonic 
degrees of freedom emerge, $n_{\bk\sigma}$ undergoes a  redistribution 
within the BZ.  For $V/4t=4$, the  probability for
occupation  of
highest momentum  states is now found in the regions around $(\pm \pi,0)$
and $(0,\pm\pi)$.
For  $\dxy$ pairing, this change in behavior of the single particle 
distribution function $n_{\bk\sigma}$ as the  chemical potential falls below the minimum of the tight binding  band is an  interesting feature.
For $s$-wave systems,  $n_{\bk\sigma}$ becomes a constant for strong coupling,
representing the Fourier  transform of a point internal wavefunction. 
Accordingly we can interpret the new structure for $n_{\bk\sigma}$ as the 
$d$-wave version of a local pair.

 One 
may speculate what the effect of this new structure 
for $n_{\bk\sigma}$ may be  at finite temperatures. Above $T_{c}$, if 
a pseudo-gap  in the normal state excitation spectrum  can arise from  
pairing fluctuations in the crossover regime \cite{Trivedi}, 
then one would expect these fluctuations  to 
 occur predominantly in the region of the BZ which has the largest 
probability of occupation by pairs.  
On the one hand, it is interesting to note from the 
bottom plot in Fig. \ref{occupation} that these regions correspond to the 
angular dependence of the  pseudo-gap in  the underdoped cuprates \cite{Marshall,Loeser}. However,  it is still unknown whether  these  materials correspond 
to such a regime.

\section{Summary}
In summary, we find that for $\dxy$ pairing, 
 only in the  dilute 
limit is it likely that a BCS-BE crossover can occur,  while it is possible 
at any 
density for $s$-wave systems. If bosonic behavior does emerge, the $\dxy$ symmetry causes the single 
particle distribution function to undergo a radical redistribution.
\section{Acknowledgments}
BCdH would like to thank  A.J. Berlinsky, M.P. Das, M.J.P. Gingras and 
C. Kallin for valuable discussions and comments. 
This work was partially funded by the 
Australian Commonwealth Government.

\end{document}